\documentclass[runningheads]{llncs}
\usepackage[T1]{fontenc}
\usepackage{graphicx}
\usepackage{float} 
\usepackage{amssymb}
\usepackage{amsmath}
\usepackage{dsfont}  
\usepackage{color}
\usepackage{subcaption}

\usepackage{tikz}
\usetikzlibrary{decorations.pathreplacing}
\usetikzlibrary{calc}

\usepackage{mathrsfs}
\usepackage{thmtools}

\usepackage{hyperref}
\usepackage{cleveref}

\begin{document}
\title{A Fair Post-Processing Method based on the MADD Metric for Predictive Student Models}
\titlerunning{A Fair Post-Processing Method based on the MADD metric}
%
\author{Mélina Verger, Chunyang Fan, Sébastien Lallé, François Bouchet, Vanda Luengo}
\authorrunning{M. Verger et al.}
%
\institute{Sorbonne Université, CNRS, LIP6, F-75005 Paris, France
\\
\email{melina.verger@lip6.fr}}
\maketitle              
\begin{abstract}
Predictive student models are increasingly used in learning environments. However, due to the rising social impact of their usage, it is now all the more important for these models to be both sufficiently accurate and fair in their predictions. To evaluate algorithmic fairness, a new metric has been developed in education, namely the \textit{Model Absolute Density Distance} (MADD). This metric enables us to measure how different a predictive model behaves regarding two groups of students, in order to quantify its algorithmic unfairness. In this paper, we thus develop a post-processing method based on this metric, that aims at improving the fairness while preserving the accuracy of relevant predictive models' results.  We experiment with our approach on the task of predicting student success in an online course, using both simulated and real-world educational data, and obtain successful results. Our source code and data are in open access
at \url{https://github.com/melinaverger/MADD}.
\end{abstract}
\section{Introduction}
\label{sec:intro}

Due to their ability to enhance educational outcomes and support stakeholders in making informed decisions, predictive student models are increasingly used in learning environments~\cite{ventura}. However, due to the rising social impact of their usage, it is thus now all the more important for these models to not only be sufficiently accurate, but also fair in their predictions. Indeed, the quality of the predictions could influence the adoption of real-world interventions and possibly produce long-term implications for students~\cite{holstein}.

To evaluate algorithmic fairness, which is broadly defined in~\cite{mehrabi}, a new metric has been recently developed, namely the \textit{Model Absolute Density Distance} (MADD)~\cite{ref_madd}. The interest of this metric is to measure how different a predictive model behaves regarding two groups, independently from its predictive performance, which will be further explained in part~\ref{subsec:motivation}. However, precisely because it is independent from predictive performance, \cite{ref_madd} recommend to use it for fairness evaluation on models that already demonstrate satisfying predictive performance to be used in real-world applications. Therefore, we would like to extend the use of this metric to unfairness mitigation as well, that is to say to improve the fairness of accurate but unfair results.

That is why, in this paper, we develop a fair post-processing method based on the MADD metric that improves the fairness of the results outputted by a model without compromising accuracy. It is important to note that this method improves the fairness of the results regarding a single chosen attribute only, since it is based on the MADD metric that evaluates fairness regarding a single attribute as well (see part~\ref{subsec:madd_def}). We then experiment with our approach on the task of predicting student success in an online course, using both simulated and real-world educational data, and for which our source code and data are in open access
at \url{https://github.com/melinaverger/MADD}. Our results show that our method successfully improves the fairness without losing the accuracy of the results. 

The remainder of this paper is organized as follows. We first describe the MADD metric in Section~\ref{sec:madd} in order to explain our MADD post-processing approach in Section~\ref{sec:postproc}. Then, we describe how we conducted our experiments in Section~\ref{sec:xp} and we report our results in Section~\ref{sec:results}. Finally, we discuss our approach in Section~\ref{sec:discussion}, and we conclude our paper in Section~\ref{sec:ccl} with future work.

\section{The \textit{Model Absolute Density Distance} (MADD) metric}
\label{sec:madd}

\subsection{Motivation of its choice}
\label{subsec:motivation}

Among the plethora of fairness metrics that have emerged from recent literature, we chose the MADD for our fair post-processing approach. The first reason is that it is a statistical metric. Indeed, fairness metrics mostly fall into three main classes, counterfactual (or causality-based), similarity-based (or individual), and statistical (or group) criteria \cite{ref_fairnessmetrics,ref_castelnovo}, but so far the metrics from the first two classes are seldom used in practice~\cite{ref_fairnessmetrics}. Using the MADD ensures an easy integration for fairness improvement in existing applications.

The second reason is that, contrary to the other statistical metrics, the MADD is able to capture the severity of the predictive errors between the different groups of students in the data (further explained in part~\ref{subsec:madd_def}). Indeed, the principle of the existing metrics consists in making comparisons of any predictive performance of a model across groups (either by independence, separation or sufficiency~\cite{kizilcec}). However, two models producing any similar predictive error proportions across groups can still exhibit very different and possibly harmful errors themselves, which is not captured by these metrics. That is why, since the MADD is precisely not based on predictive performance (then a comparison) but on the intrinsic difference with which a models behaves regarding the groups, this metric is particularly relevant for improving fairness on this distinct aspect. 

\newpage
\subsection{Preliminaries}


To provide a framework for the whole paper, let consider a binary classifier~$\mathcal{C}$ that aims at predicting student success at a course. $\mathcal{C}$ is trained on a dataset $\left\{X, Y \right\}_{i = 1}^{n}$ with $n$ the number of unique students, X the attributes characterizing the students, and $Y$ the binary target variable whose values $y_i$ can take 0 for failure and 1 for success. To apply the MADD, $\mathcal{C}$ should output both its predictions $\hat{Y}$ and the predicted probability $\hat{p_i}$ associated to each prediction: $\mathcal{C} \rightarrow 
\left\{ \hat{y_i} = \left\{0, 1\right\}, \hat{p_i} \in [0, 1] \right\} $. In the rest of the paper, we will focus on the probability related to the success prediction for every student $i$, i.e. $\hat{p_i}(\hat{y_i}=1)$, but as it is also equal to ${1 - \hat{p_i}(\hat{y_i}=0)}$, it is simpler to use the notation $\hat{p_i}$ indifferently.

Let also consider one attribute of $X$, simply noted as $a$ (instead of $x^{(a)}$ conventionally), which will be our attribute of interest, that is to say the attribute regarding which we will evaluate algorithmic fairness. This attribute $a$ should be binary, i.e. composed of two distinct groups of students, $G_0$ and $G_1$. As an example, if $a$ corresponds to having declared a disability, a student could not belong to the group of those who have not (e.g.~$G_0$) and the group of those who have (e.g.~$G_1$) declared a disability.

\subsection{Definition}
\label{subsec:madd_def}

\begin{figure}[!t]
    \centering
    \begin{subfigure}[t]{0.3\textwidth}
        \centering
        \includegraphics[width=1\textwidth]{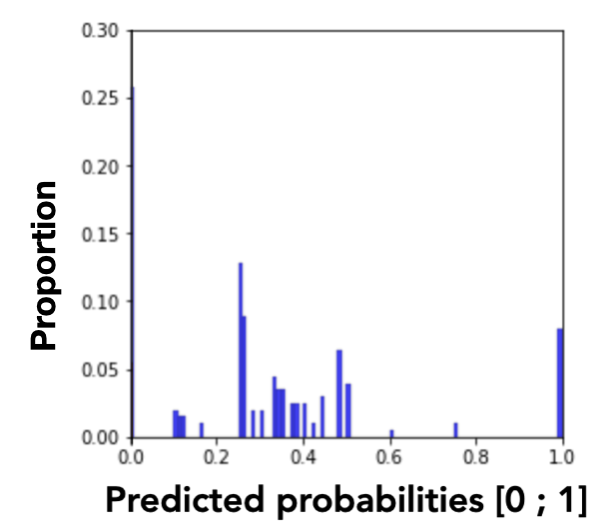}
        \caption{$D_{G_0}^a$}
        \label{fig:histoG0}
    \end{subfigure}%
    ~
    \begin{subfigure}[t]{0.3\textwidth}
        \centering
        \includegraphics[width=0.97\textwidth]{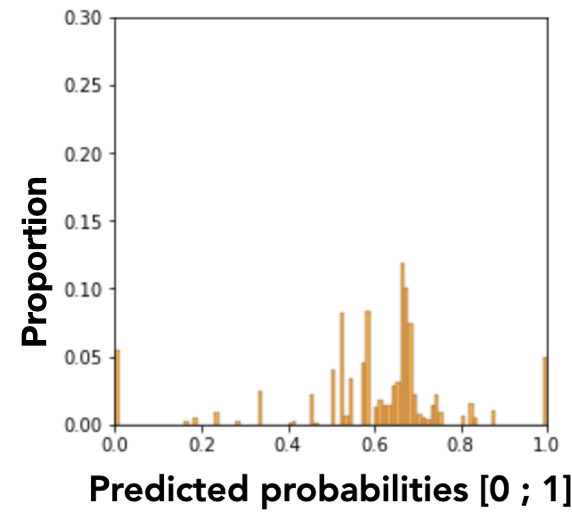}
        \caption{$D_{G_1}^a$}
        \label{fig:histoG1}
    \end{subfigure}
    ~
    \begin{subfigure}[t]{0.3\textwidth}
        \centering
        \includegraphics[width=0.98\textwidth]{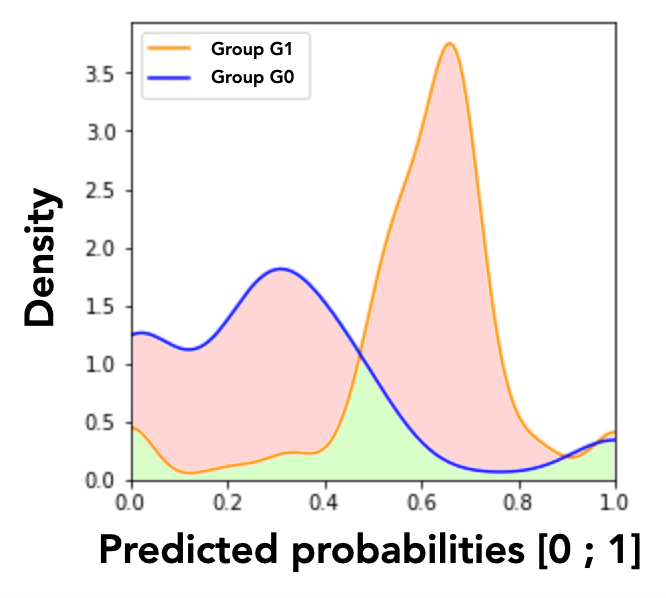}
        \caption{MADD approx.}
        \label{fig:madd_approx_visu}
    \end{subfigure}
    \caption{Representations of the MADD from~\cite{ref_madd}. (a) Proportions of predicted probabilities for group $G_0$. (b) Idem for group $G_1$. (c) Visual approximation of the MADD in the red zone, thanks to a smoothing  of the histograms (a) and (b) for easier interpretability. The smoothing has been done by kernel density estimation of the histograms~\cite{ref_madd}.}
    \label{fig:madd_expl_visu}
\end{figure}

To introduce the metric, the \textit{Model Absolute Density Distance} (MADD)~\cite{ref_madd} relies on the comparison between how a model $\mathcal{C}$ distributes its probabilities of success predictions $\hat{p_i}$ between the students of a group~$G_0$ and those of a group~$G_1$ from an attribute~$a$. To do so, the calculation of the metric needs two one-dimensional vectors of the same length, noted $D_{G_0}^a$ and $D_{G_1}^a$ and called \textit{density vectors}. They both contain the proportions of students of $G_0$ and $G_1$ respectively who receive the same predicted probabilities $\hat{p_i}$ (see Figures~\ref{fig:histoG0} and~\ref{fig:histoG1}). These proportions are noted $d_{G_0,k}^a$ and $d_{G_1,k}^a$ with $k$ the index of the distinct value of $\hat{p_i}$. The MADD is thus defined as follows~\cite{ref_madd}:
\begin{equation}
    \operatorname{MADD}\left(D_{G_0}^a, D_{G_1}^a\right) = \sum_{k=0}^m \left| d_{G_0,k}^a - d_{G_1,k}^a \right| \hspace{0.5cm} \in [ 0, 2 ]
    \label{MADDeq}
\end{equation}

To better understand what the MADD represents, a visual approximation is given in Figure~\ref{fig:madd_approx_visu} through the red zone, i.e. the zone where the two continuous density estimates (represented by continuous curves) of discrete $D_{G_0,k}^a$ and $D_{G_1,k}^a$ do not intersect. We highlight one advantage of the MADD that is to be bounded, making it objective and comparable for any data-models applications. Indeed, it is bounded between 0 and 2, and the closer the MADD is to 0, the fairer the outcome of the model is (regarding the two groups). More precisely, in the case where the MADD is equal to 0, the model produces the same probability outcomes for both groups so that ${D_{G_0}^a = D_{G_1}^a}$ and ${\text{MADD}(D_{G_0}^a, D_{G_0}^a) = 0}$. Conversely, in the most unfair case, where the model produces totally distinct probability outcomes for both groups, the MADD is equal to 2 because we directly sum over the total proportion of both groups, that is to say 1 and 1. This will give some intuition about how to improve fairness based on the MADD in the next section.

\section{The MADD Post-Processing Approach}
\label{sec:postproc}

\subsection{Purpose}

\begin{figure}[b]
    \centering
    \begin{subfigure}[t]{0.5\textwidth}
        \centering
        \includegraphics[width=0.7\linewidth]{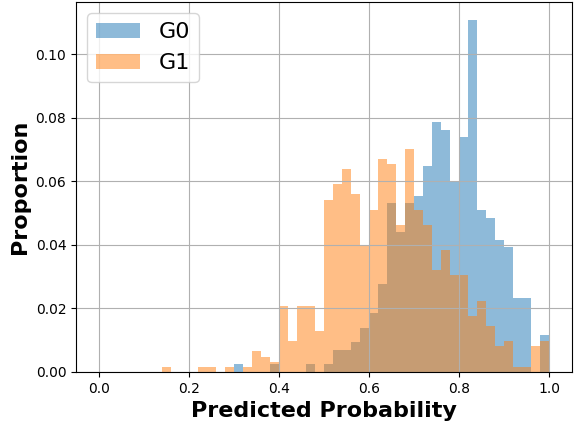}
        \caption{Before post-processing}
        \label{oulad_histo_0}
    \end{subfigure}%
    ~ 
    \begin{subfigure}[t]{0.5\textwidth}
        \centering
        \includegraphics[width=0.7\linewidth]{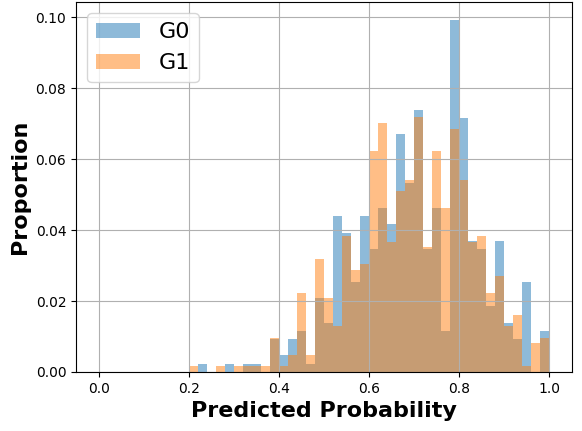}
        \caption{After post-processing}
        \label{oulad_histo_1}
    \end{subfigure}
    \caption{MADD post-processing principle. Example of two distributions of predicted probabilities (as in Fig.~\ref{fig:histoG0} and \ref{fig:histoG1}), before and after the MADD post-processing.}
    \label{oulad_histos}
\end{figure}

As introduced in the previous section, the closer the MADD is to 0, the fairer the outcome of the model is (w.r.t to attribute $a$), since the distributions of predicted probabilities are no longer distinguishable between the two groups $G_0$ and $G_1$. Thus, to illustrate how a post-processing with the MADD would work, let consider that a model tends to give higher predicted probabilities (i.e.~probabilities of success predictions) to a group, e.g.~$G_0$, than to the other, as shown in Figure~\ref{oulad_histo_0}. Therefore, the goal of the MADD post-processing is to reduce the gaps between the distributions of both groups, to obtain a result similar to what we can observe in Figure~\ref{oulad_histo_1}.

\subsection{Approach}
\label{subsec:approach}

\begin{figure}[!t]
    \centering
    \begin{subfigure}[t]{0.4\textwidth}
        \centering
        \includegraphics[width=0.8\linewidth]{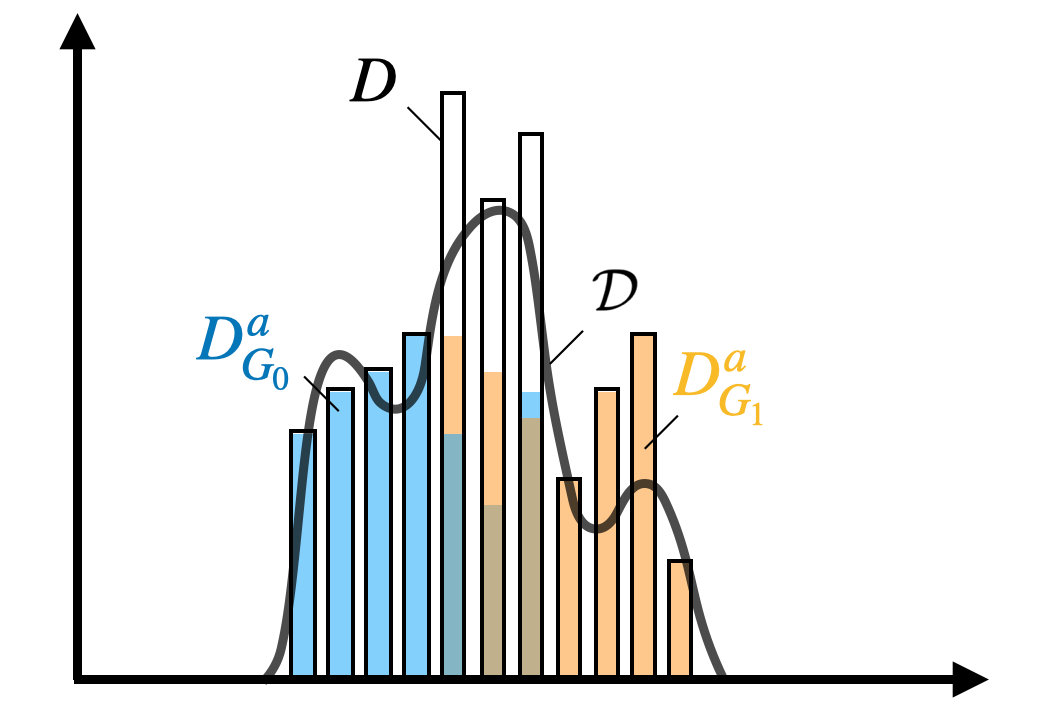}
        \caption{Principle}
        \label{principle}
    \end{subfigure}%
    ~ 
    \begin{subfigure}[t]{0.2\textwidth}
        \centering
        \includegraphics[scale=0.2]{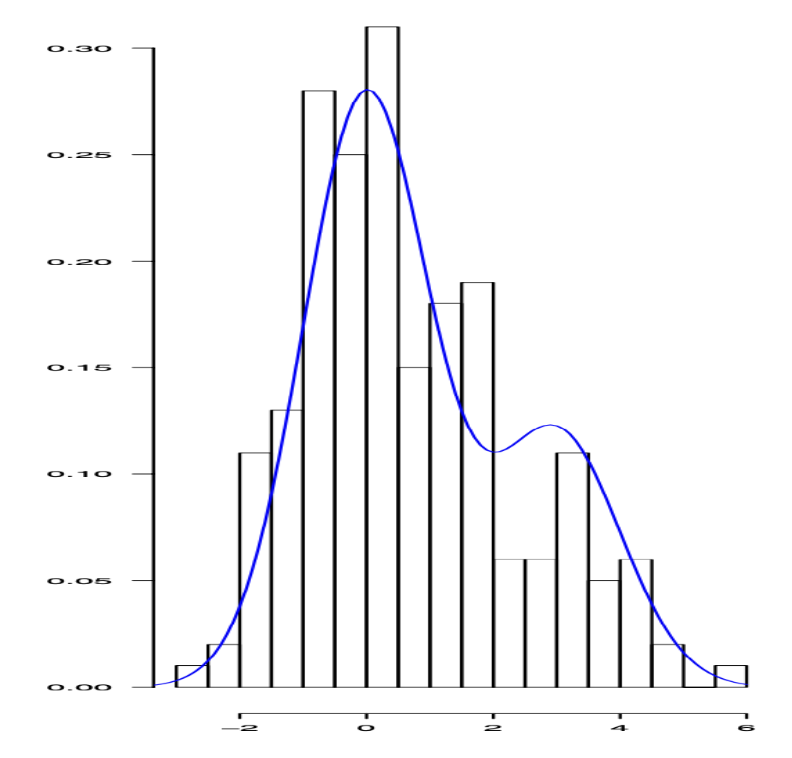}
        \caption{PDF}
        \label{fig:pdf}
    \end{subfigure}%
    ~ 
    \begin{subfigure}[t]{0.4\textwidth}
        \centering
        \includegraphics[width=0.6\linewidth]{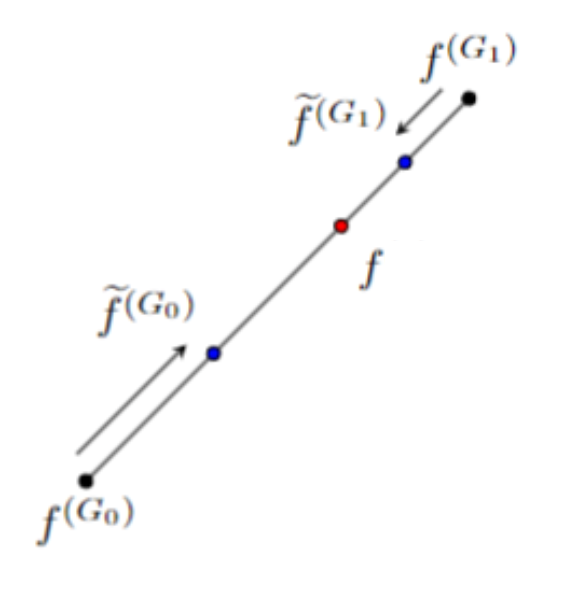}
        \caption{Influence of $\lambda$}
        \label{fig:lambda}
    \end{subfigure}
    \caption{MADD post-processing approach. (a) Illustration of the different distributions. (b) Illustration of a PDF based on an histogram. (c) Linear relationship between the PDFs (continuous space).}
    \label{fig:postproc_approach}
\end{figure}

Following up on the previous part, a question can be raised: where should the two distributions coincide? Indeed, should the distribution of $G_1$ move to the one of $G_0$ or is there a better location between the two? To solve this issue, let first note as $D$ the distribution related to all students, composed of students from both groups $G_0$ and $G_1$ (see the black histogram in Figure~\ref{principle}). In machine learning, the goal for a model is to approximate the ``true'' relationship (or prediction function ${\mathcal{X} \rightarrow \mathcal{Y}}$) between the attributes~$X$ in input and the target variable~$Y$ in output. As a consequence, we assume that a model that shows satisfying predictive performance outputs a discrete distribution $D$ which should be really close to $\mathcal{D}$, its ``true'' distribution (see the black line in Figure~\ref{principle}). Therefore, assuming having such a model, our goal is to make the distributions $D_{G_0}^a$ and $D_{G_1}^a$ coincide at the place of $D$, which should best approximate $\mathcal{D}$ (see proofs in Appendix~\ref{appendix:proofs_approach}). Indeed, this allows both to reduce the gaps between the two groups hence improve fairness and to prevent a loss in predictive performance. Therefore, the MADD post-processing is based on the following theoretical considerations.

Since $D$, $D_{G_0}^a$ and $D_{G_1}^a$ correspond to histograms, then they can be mathematically considered as discrete estimators of the ``true'' probability density functions (PDFs) they describe (see Figure~\ref{fig:pdf}) \cite{ref_histo_est}, noted as $f$, $f^{(G_0)}$ and $f^{(G_1)}$ respectively in the following. We thus want $f^{(G_0)}$ and $f^{(G_1)}$ to move towards the target $f$, as the intuition was given in the previous paragraph, but in a linear way because we want to keep the proportionality of their relative distance (see Figure~\ref{fig:lambda}), otherwise it will improve fairness more for one group than for the other. We can now define the new theoretical PDFs $\widetilde{f}$, $\widetilde{f}^{(G_0)}$ and $\widetilde{f}^{(G_1)}$ that we will estimate after the post-processing, by introducing a $\lambda$ parameter, that we call \textit{fairness coefficient of distributions convergence}, such that:
\begin{align}
\widetilde{f}^{(G_0)} = (1-\lambda) f^{(G_0)} + \lambda f \label{eq:2}\\
\widetilde{f}^{(G_1)} = (1-\lambda) f^{(G_1)} + \lambda f \label{eq:3}
\end{align}

\noindent $\lambda$ can be seen as a distance ratio (see Figure~\ref{fig:lambda}) so that ${\lambda \in [0, 1]}$, with ${\lambda = 0}$ when the PDFs of $G_0$ and $G_1$ are at their initial state and $\lambda = 1$ when they both coincide. $\lambda$ between 0 and 1 means that the distributions are getting closer (see discrete examples of distribution convergence in Figure~\ref{6histos}). The challenge is to find the highest $\lambda$ possible that best improves the fairness without affecting the accuracy of the results. However, in practice, as we do not know the true $f$, $f^{(G_0)}$ and $f^{(G_1)}$, we cannot directly compute $\widetilde{f}$, $\widetilde{f}^{(G_0)}$ and $\widetilde{f}^{(G_1)}$ as written in equations~\ref{eq:2} and~\ref{eq:3} with different values of $\lambda$. That is why we introduce \texttt{fip} in the next part.

\subsection{Implementation}
\label{subsec:implementation}


We will generate a mapping function\footnote{Here, not a mathematical function, but a programming function. See details at \url{https://github.com/melinaverger/MADD}.}, \texttt{fairness\_improved\_prediction} or in short \texttt{fip}, between the discrete estimates of $f$, $f^{(G_0)}$, $f^{(G_1)}$ (i.e. $D$, $D_{G_0}^a$, $D_{G_1}^a$) and the discrete estimates of $\widetilde{f}$, $\widetilde{f}^{(G_0)}$, $\widetilde{f}^{(G_1)}$ that we will note as $\overline{D}$, $\overline{D}_{G_0}^a$, $\overline{D}_{G_1}^a$. The purpose of \texttt{fip} is more precisely to take as inputs the $\hat{p}_i$ available at the output of a trained model and a value of $\lambda$, and to output the new fairer predicted probabilities that we note as $\overline{p}_{i}^{(\lambda)}$ (\texttt{fip}: $(\hat{p}_i$, $\lambda)$ $\mapsto$ $\overline{p}_{i}^{(\lambda)}$). Consequently, $\overline{p}_{i}^{(\lambda)}$ will allow to reconstruct the new $\overline{D}_{G_0}^a$ and $\overline{D}_{G_1}^a$, as shown in Figure~\ref{oulad_histo_1}.



\texttt{fip} will be generated as follows. Let focus on the group $G_0$ first. As we want the proportions of students having the same predicted probabilities to be kept even if the predicted probabilities values are changing with the post-processing, we will seek to make the cumulative density function (CDF) of the initial $\hat{p}_i$ of group $G_0$ being equal to the CDF of the new $\overline{p}_{i}^{(\lambda)}$ of group $G_0$. Thus, it comes that:
\begin{alignat}{2}
    & &\operatorname{CDF}_{(G_0)}(\hat{p}_{i(G_0)}) &= \overline{\operatorname{CDF}}_{(G_0)}^{(\lambda)}\left(\overline{p}_{i(G_0)}^{(\lambda)}\right) \\
    &\Longrightarrow &\overline{p}_{i(G_0)}^{(\lambda)} &=  \overline{\operatorname{CDF}}_{(G_0)}^{-1(\lambda)}(\operatorname{CDF}_{(G_0)}(\hat{p}_{i(G_0)})) 
\end{alignat}

\noindent where $\overline{\operatorname{CDF}}_{(G_0)}^{(\lambda)} = (1-\lambda)\operatorname{CDF}_{(G_0)} + \lambda \operatorname{CDF}$, and $\overline{\operatorname{CDF}}_{(G_0)}^{-1(\lambda)}$ is the general inverse function of $\overline{\operatorname{CDF}}_{(G_0)}^{(\lambda)}$. We will have the same equations for the group $G_1$. In the end, what we do is to compute the different $\operatorname{CDF}$s and $\overline{\operatorname{CDF}}$s thanks to \texttt{interp1d} and \texttt{cumtrapz} Python functions from \texttt{scipy} library that estimate their ``true'' equivalents based on the discrete values of $\hat{p}_i$ we have access to, which gives us the core of our \texttt{fip} mapping function. Now we have the ability to compute the $\overline{p}_{i}^{(\lambda)}$, let define an objective function based both on the accuracy and the fairness of the new fairer predicted probability results which depend on $\lambda$, to evaluate the outcome of our MADD post-processing method.

\subsection{Objective Function}

Similarly to existing balancing methods between accuracy and penalty values, we define the objective function as follows:

\begin{equation}
    \mathcal{L} = (1-\theta) \operatorname{AccuracyLoss}(\lambda) + \theta \operatorname{FairnessLoss}(\lambda)
\end{equation}

\noindent where $\theta \in [0, 1]$ represents the importance of the accuracy and the fairness in the objective function. Indeed, a larger $\theta$ puts more emphasis on fairness, while a smaller $\theta$ favors accuracy. The value of $\theta$ could be set by an expert depending of what one wants to put more emphasis on, or experimentally determined like what we do with $\lambda$ in part~\ref{subsec:rez_sim} for instance. The $\operatorname{AccuracyLoss}(\lambda)$, compatible with any common loss functions $\ell$ (e.g. binary cross-entropy loss), and the $\operatorname{FairnessLoss}(\lambda)$ could be defined as follows:

\begin{align}
    \operatorname{AccuracyLoss}(\lambda) &= \frac{1}{n} \sum_{i=1}^n \ell\left(\overline{p}_{i}^{(\lambda)}, y_i \right) \\
    \operatorname{FairnessLoss}(\lambda) &= \operatorname{MADD}\left(\overline{D}_{G_0}^a, \overline{D}_{G_1}^a\right)
\end{align}


Since the two losses may vary across different scales of values, one should pay particular attention to the choice of $\ell$ and the way of rescaling both losses to balance them effectively. We will show an example in part~\ref{subsec:norm_objfunc}.

\section{Experiments}
\label{sec:xp}

\subsection{Process}

\begin{figure}[!t]
    \centering
    \includegraphics[width=1\textwidth]{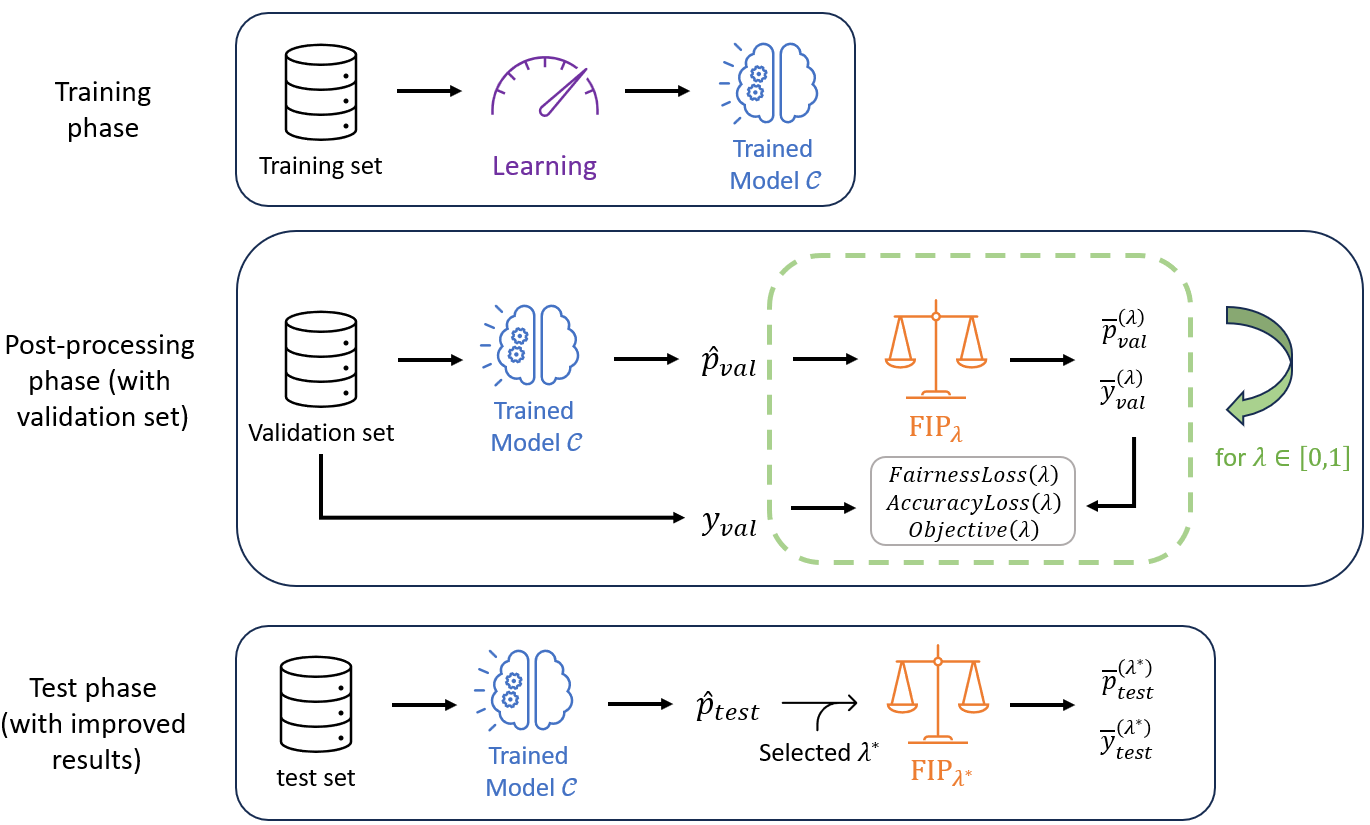}
    \caption{MADD post-processing.}
    \label{fig:process}
\end{figure}



Our MADD post-processing method, illustrated in Figure~\ref{fig:process}, can be applied for a fixed $\theta$ as follows. Let us have a training, a validation and a test sets. We first train a classifier. Then, we use this trained model on the validation set to output the predictions $\hat{y}_{i,{validation}}$ and predicted probabilities $\hat{p}_{i,{validation}}$. Next, we apply our \texttt{fip} mapping function with various values of $\lambda$ to obtain different corresponding $\overline{p}_{i,{validation}}^{(\lambda)}$. We will thus deduce the new $\overline{y}_{i,{validation}}^{(\lambda)}$ thanks to the classification threshold $t$. Now, with the new $\overline{p}_{i,{validation}}^{(\lambda)}$, $\overline{y}_{i,{validation}}^{(\lambda)}$ and the true labels $y_{i,{validation}}$, we can plot the results of our objective function depending on the $\lambda$s to find the optimal $\lambda^{*}$ that will best improve the results of the classifier. Finally, we evaluate the accuracy and the fairness of the results with the chosen $\lambda^{*}$ on the test set (i.e. with $\overline{p}_{i,{test}}^{(\lambda^*)})$, $\overline{y}_{i,{test}}^{(\lambda^{*})}$ and the true labels $y_{i,{test}}$). For the sake of simplification, in the experiments we omit \textit{training}, \textit{validation} and \textit{test} subscripts from the notations but they will be easily deduced from the context.


\subsection{Rescaled Objective Function}
\label{subsec:norm_objfunc}

%
%

\begin{align}
    \operatorname{AccuracyLoss}_{exp}(\lambda) &= \frac{1}{n} \sum_{i=1}^n \mathds{1}_{y_i \neq \overline{y}_i} \label{eq:9} \\
    \operatorname{FairnessLoss}_{exp}(\lambda) &= \frac{1}{2} \operatorname{MADD}\left(\overline{D}_{G_0}^a, \overline{D}_{G_1}^a\right) \label{eq:10}
\end{align}

\noindent For our experiments, we use the objective function $\mathcal{L}_{exp}$ composed of the rescaled terms we define as above (Equations~\ref{eq:9}\footnote{$\overline{y}_i$ corresponds to the new predictions (1 or 0) obtained thanks to the new $\overline{p}_{i}^{(\lambda)}$ thresholded with the classification threshold parameter $t$ we primarily set at 0.5 (i.e. $\overline{y}_i = 0$ when $\overline{p}_{i}^{(\lambda)} < 0.5$, $\overline{y}_i = 1$ otherwise).\label{footnote1}} and \ref{eq:10}). Both losses have thus a range of $[0, 100\%]$. Indeed, the $\operatorname{AccuracyLoss}_{exp}(\lambda)$ is the percentage of incorrect predictions, and the $\operatorname{FairnessLoss}_{exp}(\lambda)$ now represents a percentage of dissimilarity between the two distributions. Therefore, the resulting objective function $\mathcal{L}_{exp}$ is a weighted average of these two losses based on their importance. However, as a case study, we choose to give in all our experiments the same importance both to the accuracy and the fairness in the post-processing and we fix $\theta = 0.5$. Additionally, it is important to note that in the case of this $\operatorname{AccuracyLoss}_{exp}(\lambda)$, it exactly corresponds to 1 minus the standard accuracy score, which we will exploit in our results in section~\ref{sec:results}.
Our goal will be to experimentally find the optimal parameter $\lambda^*$ that minimizes this objective function $\mathcal{L}_{exp}$, with $\theta = 0.5$.

%

\subsection{Simulated Data}

%
%

To demonstrate the validity of our approach, we first experiment our MADD post-processing method on simulated data of $\hat{p}_i$ for which we know the real distributions. Simulated $\hat{p}_{i(G_0)}$ and $\hat{p}_{i(G_1)}$ are thus the predicted probabilities that we would have obtained at the output of a classifier. Let $\hat{p}_{i(G_0)}$ and $\hat{p}_{i(G_1)}$ be represented by some respective PDFs $f^{(G_0)}$ and $f^{(G_1)}$. $f^{(G_0)}$ and $f^{(G_1)}$ are respectively parts of the gamma distribution $\Gamma(4,1)$ and the normal distribution $\mathcal{N}(0.55,1)$, properly scaled along the x-axis and normalized within the interval~$[0, 1]$:

$$
\begin{alignedat}{2}
f^{(G_0)}(x) &:= \frac{1}{C_0} f_{\Gamma(4,1)} (11x) \mathbf{1}_{[0,1]}(x) &\quad &C_0:=\int_0^1 f_{\Gamma(4,1)} (11x) dx \\
f^{(G_1)}(x) &:= \frac{1}{C_1} f_{\mathcal{N}(0.55,1)} (10x) \mathbf{1}_{[0,1]}(x) &\quad &C_1:=\int_0^1 f_{\mathcal{N}(0.55,1)} (10x) dx
\end{alignedat}
$$

Based on the above PDFs, we generate $10,000$ samples of $\hat{p}_{i(G_0)}$ and $10,000$ samples of $\hat{p}_{i(G_1)}$, whose density vectors $D_{G_0}^a$ and $D_{G_1}^a$ are displayed in Figure~\ref{fig:sim_histo}. Moreover, to simulate the true label $y_i$ for each student $i$, we arbitrarily choose to pass the simulated $\hat{p}_i$ value as a parameter of a Bernoulli law: Bernoulli($\hat{p}_i$) $\in \left\{0, 1\right\}$. This will enable us to simulate how a classifier would perform before the post-processing, to compare its results with those obtained after the post-processing. The latter are thus deduced from the classification threshold parameter $t$ that we also arbitrarily set to 0.5 in this paper, and we refer the reader to the footnote~\ref{footnote1}.

\subsection{Real-world educational data}

As a second testbed for our approach, we use real-world educational data. This data, also used in~\cite{ref_madd} for fairness evaluation with the MADD, come from the Open University Learning Analytics Dataset (OULAD)~\cite{oulad} corresponding to courses offered by The Open University, a distance learning university from the United Kingdom, between 2013 and 2014. The attributes we use to predict whether a student will pass or fail a course are displayed in Table~\ref{Table_features}. The \texttt{sum\_click} attribute was the only one that was not immediately available as is, and we computed it from inner joints and aggregation on the original data. Moreover, we primarily decide in this paper to limit the data to a specific course, tagged ``BBB'' in \cite{oulad} and composed of 4,740 unique students, because it demonstrates high correlation and high imbalance w.r.t. the gender, which makes it a good candidate to analyze and improve possible algorithmic unfairness regarding this attribute. We learn a logistic classifier on this data, following a 70-15-15\% split ratio between the training, validation and test sets.


\section{Results}
\label{sec:results}

\begin{table*}[!t]
\caption{Attributes used from the OULAD~\cite{oulad}.}
\resizebox{\textwidth}{!}{%
\begin{tabular}{lll}
\hline
Name                    & Attribute type & Description                                                                                                                                                  \\ \hline
\texttt{gender}                 & binary       & the students' gender                                                                                                                                         \\
\texttt{age}                     & ordinal      & the interval of the students' age                                                                                                                    \\
\texttt{disability}              & binary      & indicates whether the students have declared a disability                                                                                                    \\
\texttt{highest\_education}      & ordinal      & the highest student education level on entry to the course                                                                                                   \\
\texttt{poverty}\footnotemark        & ordinal      & \begin{tabular}[c]{@{}l@{}}specifies the Index of Multiple Deprivation \cite{oulad} band of the place \\where the students lived during the course\end{tabular} \\

\texttt{num\_of\_prev\_attempts} & numerical      & the number of times the students have attempted the course                                                                                                   \\
\texttt{studied\_credits}        & numerical      & the total number of credits for the course the students are currently studying                                                                               \\
\texttt{sum\_click}              & numerical      & the total number of times the students interacted with the material of the course                                                                            \\ \hline
\end{tabular}
}
\label{Table_features}
\end{table*}

\footnotetext{Named as \texttt{imd\_band} in the original data.}



\subsection{Simulated Data}
\label{subsec:rez_sim}

\begin{figure}[!t]
    \centering
\includegraphics[width=\linewidth]{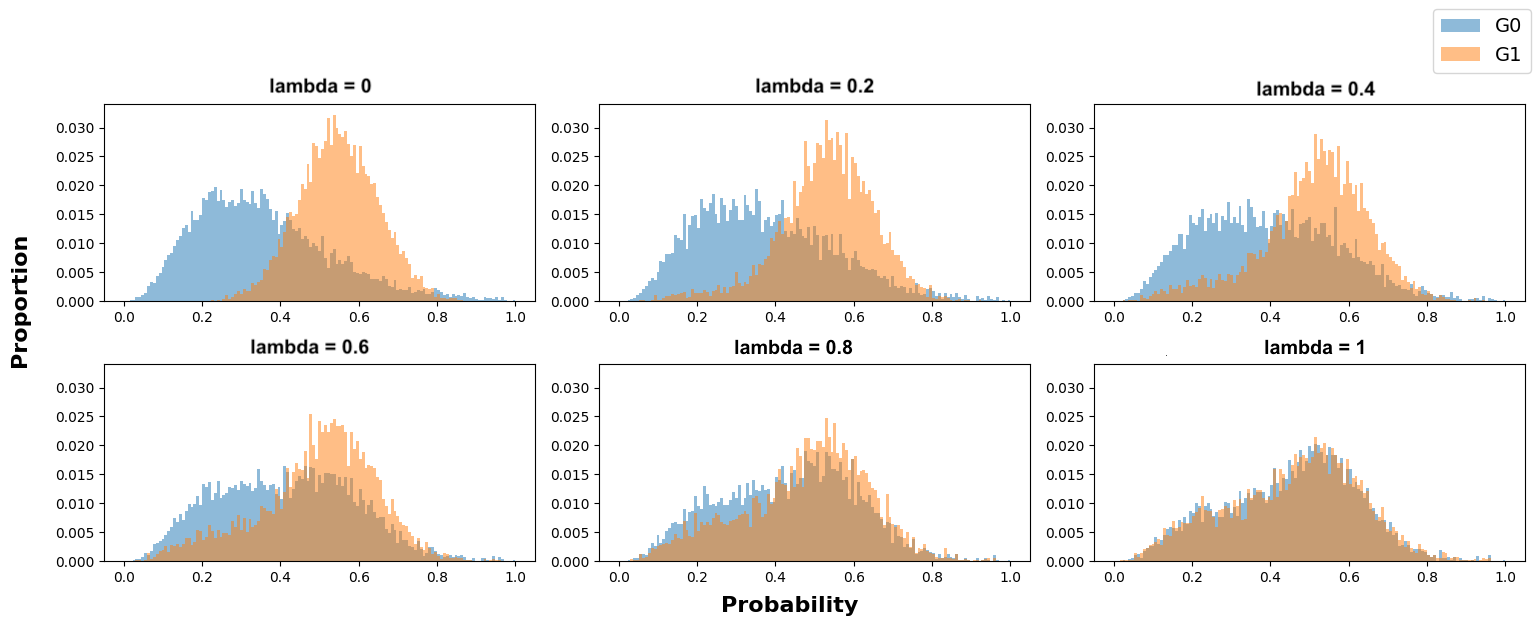}
    \caption{Effect of the MADD post-processing on the predicted probabilities with increasing values of $\lambda$.}
    \label{6histos}
\end{figure}

\begin{figure}[!ht]
    \centering
    \begin{subfigure}[t]{0.4\textwidth}
        \centering
        \includegraphics[width=1\textwidth]{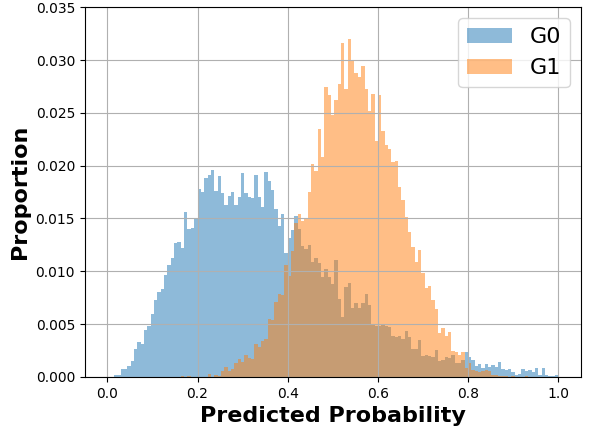}
        \caption{$D_{G_0}^a$, $D_{G_1}^a$ (i.e. $\lambda=0$)}
        \label{fig:sim_histo}
    \end{subfigure}%
    ~
    \begin{subfigure}[t]{0.4\textwidth}
        \centering
        \includegraphics[width=1\textwidth]{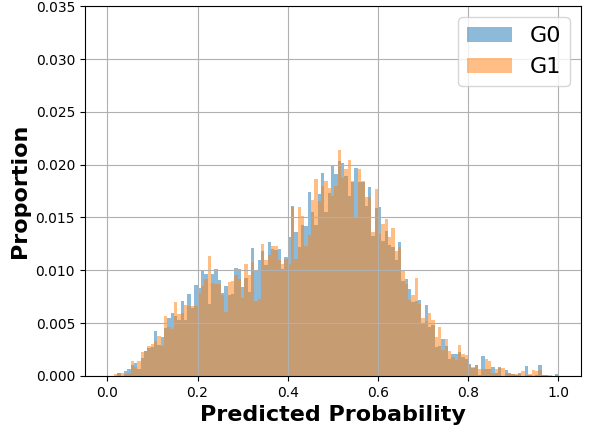}
        \caption{$\overline{D}_{G_0}^a$, $\overline{D}_{G_1}^a$ for $\lambda^{*}$}
        \label{fig:sim_rez_histo}
    \end{subfigure}
    \\
    \begin{subfigure}[t]{0.4\textwidth}
        \centering
        \includegraphics[width=1\textwidth]{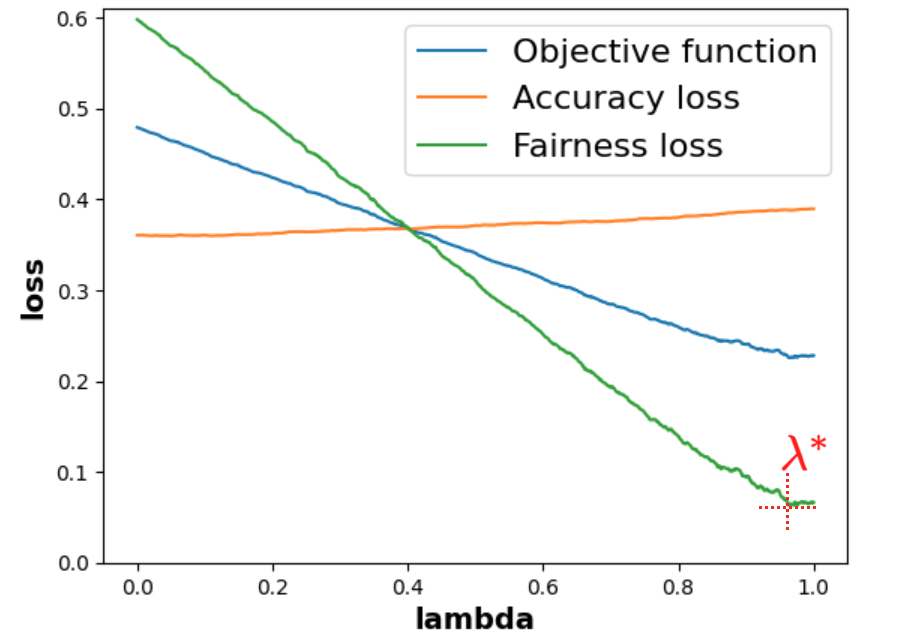}
        \caption{First simulation}
        \label{fig:sim_losses}
    \end{subfigure}
    ~
    \begin{subfigure}[t]{0.4\textwidth}
        \centering
        \includegraphics[width=1\textwidth]{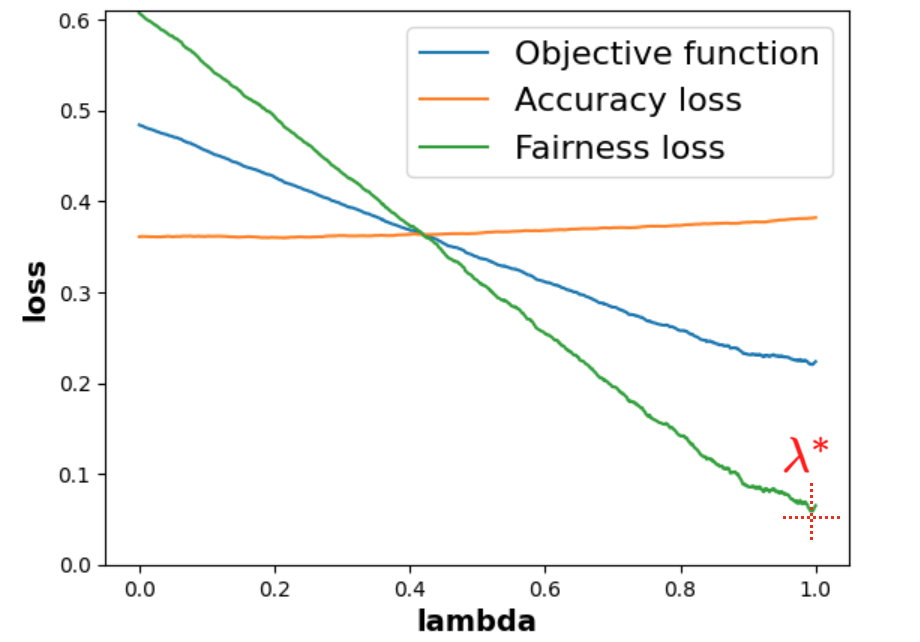}
        \caption{Additional simulation}
        \label{fig:sim_losses2}
    \end{subfigure}
    \caption{Simulated data results. (a) Histograms of $D_{G_0}^a$ and $D_{G_1}^a$ from simulated $\hat{p}_{i(G_0)}$ and $\hat{p}_{i(G_1)}$.  (b) Histograms of the new $\overline{D}_{G_0}^a$ and $\overline{D}_{G_1}^a$. (c, d) Objective function (total loss), accuracy loss and fairness loss.}
    \label{fig:sim_rez}
\end{figure}



We generate 1,000 values of $\lambda$ with a constant step in its interval $[0,1]$, and we compute all the corresponding $\overline{p}_i(\lambda)$, to obtain the relationships between the next three $\mathcal{L}_{exp}$, $\operatorname{AccuracyLoss}_{exp}(\lambda)$, $\operatorname{FairnessLoss}_{exp}(\lambda)$ and $\lambda$. In Figure~\ref{6histos}, we present how the new predicted probabilities progress with some increasing values of $\lambda$. In Figure~\ref{fig:sim_losses}, we display for all values of $\lambda$ the evolution of $\mathcal{L}_{exp}$, $\operatorname{AccuracyLoss}_{exp}(\lambda)$ and $\operatorname{FairnessLoss}_{exp}(\lambda)$. We remind that we set $\theta = 0.5$ as we decided to give an equal importance to both the accuracy and the fairness in the post-processing. As we can see (Fig.~\ref{fig:sim_losses}), on the one hand, when $\lambda$ increases the accuracy loss increases too (while we want to minimize it), but only slightly ($0.361$ to $0.390$ i.e. about +8\%). On the other hand, the fairness loss, which corresponds to half of the MADD, significantly drops as what we look for ($0.598$ to its lowest at $0.063$, i.e. about -90\%). In addition, the objective function $\mathcal{L}_{exp}$ reaches its minimum value $0.226$ at $\lambda^{*} = 0.970$, almost 1. Therefore, if we accept to loose about 8\% of accuracy (we can make this interpretation because of how we defined our $\operatorname{AccuracyLoss}_{exp}(\lambda)$), then by choosing $\lambda^{*} = 0.970$ we would increase the fairness of the results by 90\% w.r.t the MADD criterion. After that, we only have to pass $\lambda^{*} = 0.970$ and the $\hat{p}_i$ as inputs of \texttt{fip} to obtain our new fairer predicted probabilities as shown in Figure~\ref{fig:sim_rez_histo}. We repeate the simulation by generating some other random 10,000 samples for each group, and the results are very similar (see Figure~\ref{fig:sim_losses2}), which strengthens the estimation of $\lambda^{*}$ being close to 1.
To conclude, this experiment, based on a simulated and ideal case study with sufficient data, demonstrates that the MADD post-processing manages to preserve a reasonably similar level of accuracy while significantly improving the fairness of the results. Let us now apply it with real-world data in the next part.





\begin{figure}[!ht]
    \centering
    \begin{subfigure}[t]{0.5\textwidth}
        \centering
        \includegraphics[width=.85\linewidth]{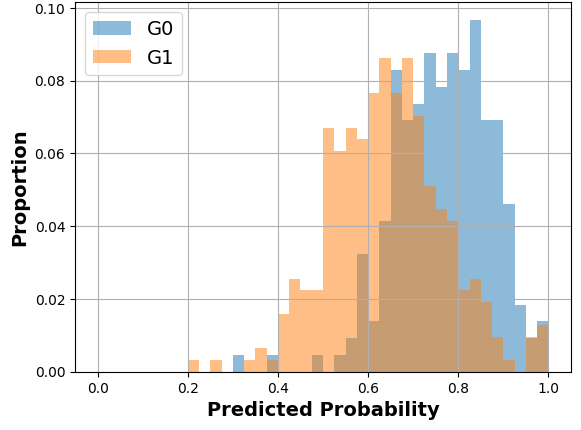}
        \caption{$D_{G_0}^a$, $D_{G_1}^a$ (i.e. $\lambda=0$)}
        \label{fig:histo_init_oulad}
    \end{subfigure}%
    ~
    \begin{subfigure}[t]{0.5\textwidth}
        \centering
        \includegraphics[width=.85\linewidth]{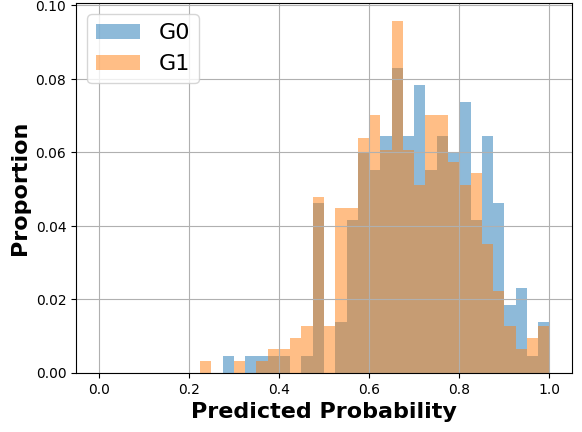}
        \caption{$\overline{D}_{G_0}^a$, $\overline{D}_{G_1}^a$ for $\lambda^{*}$}
        \label{fig:histo_final_oulad}
    \end{subfigure}%
    \\
    \begin{subfigure}[t]{0.5\textwidth}
        \centering
        \includegraphics[width=.85\linewidth]{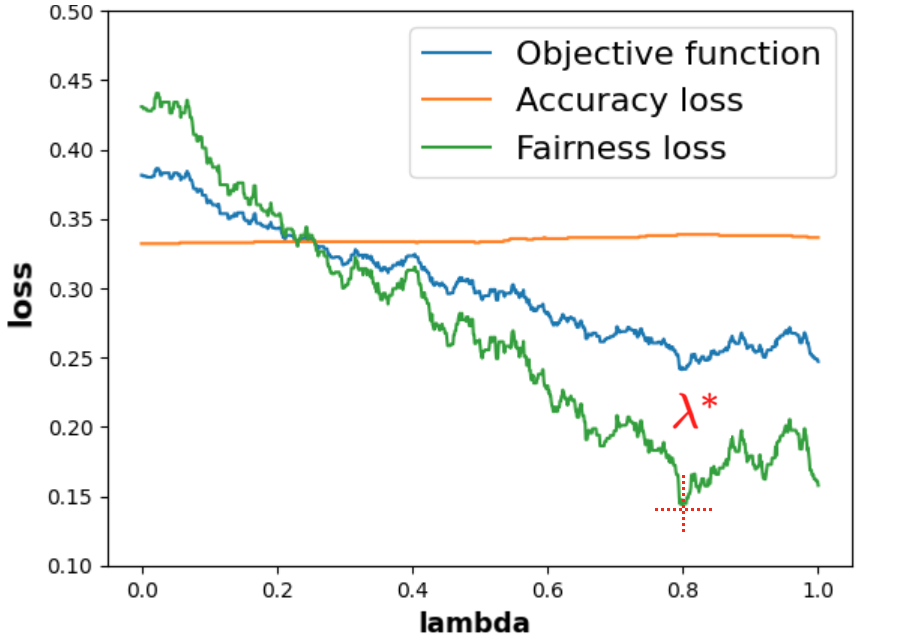}
        \caption{Validation set}
        \label{fig:2Dval_oulad}
    \end{subfigure}%
    \caption{Real-world educational data results.}
    \label{real_2D}
\end{figure}

\subsection{OULAD Data}

Again, we display in Figure~\ref{fig:2Dval_oulad} the evolution of $\mathcal{L}_{exp}$, $\operatorname{AccuracyLoss}_{exp}(\lambda)$ and $\operatorname{FairnessLoss}_{exp}(\lambda)$, for the values of $\lambda$ we generated in part~\ref{subsec:rez_sim}.
When $\lambda$ increases, the accuracy loss remains almost constant ($0.332$ to $0.336$ i.e. about +1\%), and the fairness loss significantly drops again ($0.431$ to its lowest at $0.158$, i.e. about ${-63\%}$). However, the Figure~\ref{fig:2Dval_oulad} shows a lot more variability than in the previous case study, which comes from the much lower number of samples in the validation set (about 700). This loss in precision makes more challenging to find an optimal $\lambda^{*}$. Indeed, we see that the minimum of the objective function (0.241)
is not necessarily reached at the optimal $\lambda^{*}$ (0.798) because the $\operatorname{FairnessLoss}(\lambda)$ seems to keep decreasing but the computation is not precise enough. However, let choose this $\lambda^{*} = 0.798$ anyway to visually evaluate how close or far the results are from satisfying fairness. We can still observe satisfying results from Figures~\ref{fig:histo_init_oulad} to~\ref{fig:histo_final_oulad}. To conclude, similarly to machine learning, the MADD post-processing is sensitive to the number of data, but it still show very successful fairness improvement without losing the accuracy of the results.


%

\section{Discussion}
\label{sec:discussion}

Both experiments on simulated data and real-world data show that a post-processing based on the MADD metric successfully improves fairness while preserving a reasonably constant accuracy. The strength of our approach lies precisely on having found, for every possible value of $\lambda$, i.e. the \textit{fairness coefficient of distributions convergence}, how to compute the two new predicted probability distributions to make them fairer and to keep the accuracy preserved. 

Indeed, the slight change in accuracy can be explained as follows. Since the model predicts success (i.e.~1) for all students whose predicted probabilities exceed the threshold~$t$, then after the post-processing only those whose new predicted probabilities pass the threshold~$t$ affect the accuracy (i.e. passing from 0 to 1 in their prediction). More precisely, we found that the accuracy loss is a function of the quantile $\overline{\operatorname{CDF}}^{(\lambda)}(t)$, and the closer $\overline{\operatorname{CDF}}^{(\lambda)}(t)$ is to the quantile $\operatorname{CDF}(t)$ of the original distribution, the smaller the accuracy loss is, and vice versa. Thus, the selection of the overall distribution as the target for the convergence of the distributions of the two new probabilities makes the change in $\overline{\operatorname{CDF}}^{(\lambda)}(t)$ small compared to the original distribution, and therefore ensures a small loss of accuracy. Besides, it is important to note the role of~$t$ on the students whose prediction changes.

In addition, from the linear relationship the MADD post-processing is based on, it guarantees that the $\operatorname{FairnessLoss}(\lambda)$ will always linearly decrease with $\lambda$. Indeed, the $\operatorname{FairnessLoss}(\lambda)$ is twice the MADD of the new predicted probabilities, that are getting linearly closer with $\lambda$. Thus, as the MADD linearly decreases while they linearly get closer (because of how the MADD is defined), so do the $\operatorname{FairnessLoss}(\lambda)$. 

However, a limitation of our approach is that even if ${\lambda = 1}$, the new generated distributions $\overline{D}_{G_0}^a$ and $\overline{D}_{G_1}^a$ will not overlap exactly, but will be infinitely close as the sample size increases. Indeed, it comes from the fact that the new predicted probabilities are computed via a histogram-based CDF, and the speed of convergence of the histogram as an estimator also depends on the sample size.






\section{Conclusion}
\label{sec:ccl}

In this paper, we present a fair post-processing method based on the MADD metric. We apply this method to the task of predicting student success at the course level, with the code and data in open access at \url{https://github.com/melinaverger/MADD}. This post-processing method allows to improve the fairness of models' results deemed sufficiently accurate but unfair. It does not require to have access to the original data nor the trained model itself. Finally, experimenting with various values of $\theta$ is part of our future work, as well as extending our approach to the consideration of multiple attributes for a more global fairness improvement.






\appendix
\section{Proofs of \ref{subsec:approach}}
\label{appendix:proofs_approach}

We set $C$ to be a random variable with probability density function $\mathcal{D}$, representing the predicted probability value of the output of the model $\mathcal{C}$, and $S$ to be a random variable subject to Bernoulli distribution, representing the value of the sensitive parameter $a$. Thus, $\mathcal{D}^{G_0}$ and $\mathcal{D}^{G_1}$ are the probability density functions of the conditional distributions $C | S=0$ and $C | S=1$, respectively. According to the law of total probability, we have:
$$
\begin{alignedat}{2}
&\quad &\mathbb{P}\left( C \leq t \right) &= \mathbb{P}\left( C \leq t \mid S=0 \right) \ \mathbb{P}\left( S=0 \right) + \mathbb{P}\left( C \leq t \mid S=1 \right) \ \mathbb{P}\left( S=1 \right) \\
&\Longleftrightarrow & F(t) &= F^{G_0}(t) \ \mathbb{P}\left( S=0 \right) + F^{G_1}(t) \ \mathbb{P}\left( S=1 \right) \\
&\Longleftrightarrow & \mathcal{D}(t) &= \mathcal{D}^{G_0}(t) \ \mathbb{P}\left( S=0 \right) + \mathcal{D}^{G_1}(t) \ \mathbb{P}\left( S=1 \right)
\end{alignedat}
$$
Where $F, F^{G_0}, F^{G_1}$ are the cumulative distribution functions (CDFs) of $\mathcal{D}$, $\mathcal{D}^{G_0}$, $\mathcal{D}^{G_1}$, respectively. Since $\mathbb{P}\left( S=0 \right) + \mathbb{P}\left( S=1 \right) = 1$, $f$ is a linear combination of $\mathcal{D}^{G_0}$ and $\mathcal{D}^{G_1}$, and $\mathcal{D}$ lies between $\mathcal{D}^{G_0}$ and $\mathcal{D}^{G_1}$ in the function space (i.e., $\mathcal{D}^{G_0}, \mathcal{D}, \mathcal{D}^{G_0}$ are collinear.

This property is also true for estimators obtained from observed values. In fact, the definition of the sequence of the heights of the histogram is: for the $m$ equal sub-intervals $\left[ \frac{k-1}{m}, \frac{k}{m} \right]$ for all $k \in \{1, \ldots, m\}$ on $[0,1]$,

\begin{align*}
&D_{G_0}^a 
&:= \left\{ d_{G_0, k} \mid \forall k \in \{ 1,\ldots, m \} \right\}, 
&\text{ with } d_{G_0, k} 
&:= &\frac{N_{G_0, k}}{n_0} 
&:= &\frac{1}{n_0}\sum_{i \in G_0} \mathbf{1}_{\hat{p}_i \in I_k} \\
&D_{G_1}^a 
&:= \left\{ d_{G_1, k} \mid \forall k \in \{ 1,\ldots, m \} \right\}, &\text{ with } d_{G_1, k} 
&:= &\frac{N_{G_1, k}}{n_1} 
&:= &\frac{1}{n_1}\sum_{i \in G_1} \mathbf{1}_{\hat{p}_i \in I_k} \\
&D_{G}
&:= \left\{ d_{G, k} \mid \forall k \in \{ 1,\ldots, m \} \right\}, 
&\text{ with } d_{G, k} 
&:= &\frac{N_{G_0, k} + N_{G_1, k}}{n_0 + n_1} 
&:= &\frac{1}{n_0 + n_1}\sum_{i \in G} \mathbf{1}_{\hat{p}_i \in I_k}
\end{align*}

And because for all $k \in \{1,\ldots, m\}$, we have:

\begin{align*}
d_{G, k} &= \frac{N_{G_0, k} + N_{G_1, k}}{n_0 + n_1} = \frac{n_0 \  d_{G_0,k} + n_1 \  d_{G_1,k}}{n_0 + n_1} \\
&= \frac{n_0}{n_0 + n_1} d_{G_0,k} + \frac{n_1}{n_0 + n_1} d_{G_1,k}
\end{align*}

Also, $f^{(G_0)}, f, f^{(G_1)}$ are based on $D_{G_0}^a, D_{G}, D_{G_1}^a$, respectively:

\begin{align*}
    f^{(G_0)}(x) &:= \sum_{k=i}^m d_{G_0, k} \mathbf{1}_{x \in I_k} \\
    f^{(G_1)}(x) &:= \sum_{k=i}^m d_{G_1, k} \mathbf{1}_{x \in I_k} \\
    f(x) &:= \sum_{k=i}^m d_{G, k} \mathbf{1}_{x \in I_k}
\end{align*}

Therefore, $f^{(G)}(x) = \frac{n_0}{n_0 + n_1} f^{(G_0)}(x) + \frac{n_1}{n_0 + n_1} f^{(G_1)}(x)$, so $f^{(G_0)}, f, f^{(G_1)}$ are also collinear (see Figure \ref{fig:lambda}). This is not a coincidence; in fact, as histogram estimators, when $(n_0, n_1) \to +\infty$, $\left( f^{(G_0)}, f, f^{(G_1)} \right) \to \left( \mathcal{D}_{G_0}^a, \mathcal{D}_{G}, \mathcal{D}_{G_1}^a \right)$.

\section{Proof of \ref{subsec:implementation}}
\label{appendix:proofs_CDF}

According to Inverse transform sampling \cite{ref_inverse}, we have the following two theorems:

\begin{theorem}\label{th1}
    Let $\mathcal{A}$ be a distribution and $F_{\mathcal{A}}$ be the cumulative distribution function of that distribution. If $X$ obeys the distribution $\mathcal{A}$ i.e. $X \sim \mathcal{A}$, then $F_{\mathcal{A}}(X) \sim \mathcal{U}_{[0, 1]}$, where $\mathcal{U}_{[0,1]}$ is a uniform distribution over $[0,1]$.
\end{theorem}

\begin{theorem}\label{th2}
    Let $U \sim \mathcal{U}_{[0,1]}$ and $F^{-1}_{\mathcal{A}}$ be the generalised inverse function of $F_{\mathcal{A}}$, then $F^{-1}_{\mathcal{A}}(U) \sim \mathcal{A}$.
\end{theorem}

Take $G_0$ as an example. By definition, the newly generated prediction $\overline{p}_{i}^{(\lambda)}$ is $\overline{\operatorname{CDF}}_{(G_0)}^{-1(\lambda)}(\operatorname{CDF}_{(G_0)}(\hat{p}_i))$, and by Theorem \ref{th1}, we have $\operatorname{CDF}_{(G_0)}(\hat{p}_i) \sim \mathcal{U}_{[0,1]}$, so $\overline{\operatorname{CDF}}_{(G_0)}^{-1(\lambda)}(\operatorname{CDF}_{(G_0)}(\hat{p}_i))$ obeys the newly generated distribution according to Theorem \ref{th2}. Furthermore, since the CDF is monotone increasing and the inverse function does not change the monotonicity, $\overline{\operatorname{CDF}}_{(G_0)}^{-1(\lambda)}$ is also monotone increasing, which means that $\forall i,j, \hat{p}_i \geq \hat{p}_j \Longrightarrow \overline{p}_{i}^{(\lambda)} \geq \overline{p}_{i}^{(\lambda)}$. The conclusion on $G_1$ follows in the same way.

%
%
%
\bibliographystyle{splncs04.bst}
\bibliography{MADDpp.bib}
\end{document}